\documentstyle{mn}
 
\input psfig.sty

\title[GM~Auriagae's Proptoplanetary Disc]
     {Constraints on a planetary origin for the gap in GM Aurigae's protoplanetary disc}

\author[Rice et al.]{W.K.M.~Rice$^1$, Kenneth Wood$^1$, 
P.J.~Armitage$^{2,3}$, B.A.~Whitney$^4$, J.E.Bjorkman$^5$\\
$^1$School of Physics \& Astronomy, University of St Andrews,
North Haugh, St Andrews, Kingdom of Fife, KY16 9SS, Scotland\\
$^2$JILA, Campus Box 440, University of Colorado, Boulder, CO 80309-0440, USA\\
$^3$Department of Astrophysical and Planetary Sciences, Boulder CO 80309-0391, USA\\
$^4$Space Science Institute, 3100 Marine Street, Suite A353, Boulder, CO 80303, USA\\
$^5$Ritter Observatory, Department of Physics \& Astronomy, University of Toledo,
Toledo, OH 43606, USA}

\begin{document}

\maketitle

\begin{abstract}
The unusual spectral energy distribution (SED) of the classical T Tauri star GM Aurigae
provides evidence for the presence of an inner disc hole extending to several au. Using a 
combination of hydrodynamical simulations and Monte Carlo radiative transport, we investigate
whether the observed SED is consistent with the inner hole being created and maintained by
an orbiting planet. We show that an $\sim 2 M_{\rm Jupiter}$ planet, orbiting at 2.5~au in a 
disc with mass $0.047 M_\odot$ and radius 300~au, provides a good match both to the SED
and to CO observations which constrain the velocity field in the disc. A range of planet
masses is allowed by current data, but could in principle be distinguished with further 
observations between $3$ and $\sim 20$ microns. Future high precision astrometric 
instruments should also be able to detect the motion of the central star due to an orbiting
Jupiter mass planet. We argue that the small number of T Tauri stars with SEDs resembling
that of GM~Aur is broadly consistent with the expected statistics of embedded migrating planets. 
\end{abstract}


\section{Introduction}

The detection of extrasolar planets and planetary systems (e.g., Mayor
\& Queloz 1995) has 
invigorated studies of planetary formation (Lissauer 1993; Boss 1998, 2000), 
migration (e.g., Lin \& Papaloizou 1986; Lin et al. 1996; Trilling et al. 1998;
R. Nelson et al. 2000), 
and 
interaction with a protoplanetary disc (Lin \& Papalizou 1979a, b; Artymowicz 
\& Lubow 1994).  Recent theoretical 
work has investigated mechanisms for halting planetary migration and 
explaining the statistical distribution of the orbital radii of extrasolar 
planets (Armitage et al. 2002; Kuchner \& Lecar 2002; Trilling, Lunine \& Benz 2002). 
Theoretical models for the interactions of (proto)planets with a protoplanetary 
disc show that large low density regions (gaps) can be created within 
the disc (Lin \& Papaloizou 1979a, b; Artymowicz \& Lubow 1994).  
The detection of these gaps would provide strong 
support for planetary formation theories, but current observations 
do not have the spatial resolution to directly image the au sized gaps 
that are predicted.  Future instrumentation, such as ALMA, may allow 
direct detection of gaps (Wolf, Henning \& Kley 2002), but for now we are restricted to 
spectral techniques for inferring the presence of disc gaps.  

Spectral 
energy distributions (SEDs) from protoplanetary discs arise from 
the thermal reprocessing of starlight and dissipation of viscous 
accretion luminosity (Lynden-Bell \& Pringle 1974; Adams, Lada \& Shu 1987;
Kenyon \& Hartmann 1987).  
The resulting SEDs display characteristic infrared 
excesses that are sensitive to the size, shape, and mass of the disc.  
Disc gaps may manifest themselves in SEDs by removing material that 
would normally have contributed to the SED at a given wavelength:  gaps 
close to the star result in smaller than usual near-IR excesses, while 
gaps at large disc radii will cause distortions to the SED at longer 
wavelengths (Beckwith 1999).  The classical T~Tauri stars GM~Aur and 
TW~Hya have SEDs that indicate the inner regions of their discs have been cleared 
of material out to around 4~au from the central stars (Koerner, Sargent, 
\& Beckwith 1993; Chiang \& Goldreich 1999; Schneider et al. 2002; 
Calvet et al. 2002). 

There are a number of possible mechanisms for clearling such
gaps. Photoevaporation by ultraviolet radiation, either from the central star or
from
surrounding radiation, can produce a disc wind at radii where the sound speed 
exceeds the escape velocity (Clarke et al. 2001). 
The inner regions of the disc will then be cleared on a viscous timescale 
once the wind mass loss rate is comparable to the accretion rate. This, however,
generally requires an accretion rate two orders of magnitude lower than the $10^{-8}
M_{\odot} \rm{yr}^{-1}$ accretion rate expected
for GM~Aur (Gullbring et al. 1998). Clarke et al. (2001) also assumed
a constant ionizing flux. Matsuyama et al. (2002) show that the clearing
of the inner disc is even less likely if the decrease
in ionizing flux with decreasing accretion rate is included in the calculation.
A. Nelson et al. (2000) also suggest that
viscous dissipation and the dissipation of gravitational instabilities may
increase the midplane temperature of the inner regions of the disc to above the
grain destruction temperature. This would then manifest itself as a reduction in
the flux of radiation with wavelengths between a few and ten microns. Their
simulation, however, assumed a disc luminosity ($0.5 L_{\odot}$) 
an order of magnitude greater 
than that calculated by Gullbring et al. (1998) for 
GM~Aur ($0.071 L_{\odot}$). 

It has been suggested (Calvet et al. 2002) that the inner regions of TW Hydra's
disc could be
cleared by tidal interactions with an embedded planet (Lin \& Papalizou 1979a, b). 
In this paper we consider if the gap in GM~Aur's disc could be formed in the
same way. We use a dynamical model of an approximately Jupiter
mass planet
in a protoplanetary accretion disc, with parameters appropriate for GM~Aur 
(e.g., Koerner, Sargent \& Beckwith 1993; 
Simon, Dutrey \& Guilloteau 2000; Schneider et al. 2002), 
to investigate if the resulting disc 
structure can reproduce the observed SED. We also consider 
other planet masses to test if the SED can be used to
constrain the planet mass. Estimates for GM~Aur's age vary from
$\sim 1.5$ Myr (Beckwith et al. 1990; 
Siess, Forestini \& Bertout 1999) to $\sim 10$ Myr 
(Simon \& Prato 1995; Hartmann 2002). There has therefore been sufficient time for an embedded planet 
to have formed via core accretion (Lissauer 1993) or via
gravitational collapse (Boss 1998, 2000). The observation of a much younger system 
(e.g., $< 1$ Myr)
with an SED signature indicating the presence of a disc gap would, if the gap could be shown
to be due to the presence of an embedded planet,
suggest that either core accretion occurs on a much
shorter timescale than expected or that planet formation can occur via an alternative 
mechanism such as gravitational collapse. In section~2 we describe and present the results of 
our dynamical simulations, in \S~3 we compare model SEDs with the observed SED of GM~Aur, and 
we summarize our results in \S~4.

\section{Dynamical Simulations}

\subsection{Dynamical Model and Initial conditions}
The simulations presented here were performed using smoothed particle hydrodynamics
(SPH), a Lagrangian hydrodynamics code (e.g., Benz 1990; Monaghan 1992). We initially
assume a planet with mass $M_p = 1.7 M_J$ orbits a star, 
with mass
$M_* = 0.85 M_{\odot}$, at a radius of $r_p = 2.5$ au. Both the 
central star and planet are modelled as point masses
onto which gas particles can accrete if they approach to within the sink radius
(Bate et al. 1995).  The planet orbits within a circumstellar disc of 
mass $M_{\rm disc} = 0.047 M_{\odot}$, modelled using 300000 SPH gas particles distributed 
such as to give a surface density profile of $\Sigma \propto r^{-1}$ (D'Alessio et al.
1998). The disc extends from $r_{\rm in} = 0.25$ au to $r_{\rm out} = 300$ au. The 
disc temperature varies as $T \propto r^{-1/2}$, as expected for discs heated
by starlight (D'Alessio et al. 1998), and we assume hydrostatic 
equilibrium in the vertical direction, giving a central density 
profile of $\rho \propto r^{-2.25}$.  For the SPH simulation, 
the disc is assumed to be locally isothermal
and the above temperature profile is maintained throughout the simulation. The
parameters chosen are appropriate for GM~Aur based on calculations of the stellar
mass (Simon, Dutrey \& Guilloteau 2000) and the disc mass and size (Schneider et al.
2002).

The vertical scale height, $H$, 
of the disc is given by $H^2=c_s^2 r^3 / G M_*$ where $c_s$ is the 
local sound speed, $r$ is the radius, $G$ is the gravitational constant, 
and $M_*$ is the stellar mass. The temperature is normalised such as to give
a scale height of 10.5 au at 100 au. The scale height at 2.5 au is then 0.1 au, giving
an $H/r$ value of 0.04. This choice of scale height was based on an SED model
using an analytic disc geometry 
and is similar to that obtained by Schneider et al.(2002). In using the above form
for the scale height, we assume that the dust has the same scale height as the gas. Although
this may not necessarily be the case, the need to use a dust scale height that has the
same form as a standard equilibrium disc (e.g., Pringle 1981) would seem to make this a 
reasonable assumption.

Using 300000 SPH particles the vertical profile 
of the disk can be modelled accurately within $\sim 3$ 
scale heights of the disc midplane. For
higher altitudes, the density is calculated analytically using (e.g., Pringle 1981)

\begin{equation}
\rho(z,r) = \rho_c(r)\mathrm{exp}(-z^2/2 H^2)
\label{zdens}
\end{equation}

\noindent where $z$ is the height above the disc midplane, and $\rho_c(r)$ is the central 
density a distance $r$ from the central star, determined using the dynamical model. 
Very little mass is contained in
these regions of the disc and is hence unimportant for the dynamical calculation. 
It is, however, extremely important for calculating the radiative transfer 
through the disc (see D'Alessio et al. 1998).

Self-gravity is included in the calculation and both the point masses and
the gas use a tree to calculate gravitational forces (Benz et al. 1990). The
inclusion of self-gravity means that the central
star and planet are both free to move under the gravitational influence of the 
disc.  A great saving in 
computational time is obtained by using individual, particle time-steps (Bate et al. 
1995; Navarro \& White 1993) which are limited by the Courant condition and a
force condition (Benz et al. 1990).

\subsection{Planet-disc interactions}

A planet orbiting within a circumstellar disc will tidally interact with the disc.
Disc material exterior to the planet will gain angular momentum
from the planet and will move to larger radii, 
while material interior to the planet will lose angular momentum and move to smaller
radii
(Lin \& Papaloizou 1979a, b; Artymowicz \& Lubow 1994). Viscous processes within the
disc will tend to act against the tidal forces and the formation of a gap will depend on
the tidal forces overcoming these viscous processes (Lin \& Papaloizou
1979a). This depends on the relative magnitudes of the 
mass ratio of the planet and the star, $q$, and the Reynolds number within the disc,
${\mathcal R}$. If $q > {\mathcal R}^{-1}$ a gap will form, while if 
$q^2 > {\mathcal R}^{-1}$
the disc will be truncated at the outer Lindblad resonance. The Reynolds number
is given by $\Omega r^2 / \nu$ where $\Omega$ is the angular frequency, $r$ is the radius,
and $\nu$ is the turbulent viscosity in the disc generally described as an
alpha viscosity (Shakura \& Sunyaev 1973).

For the initial simulation ($M_p = 1.7 M_J$, $M_{\rm disc} = 0.047 M_{\odot}$)
considered here $q = 0.002$, and the Reynolds number, determined from
the artificial viscosity included in the code (Murray 1996), is $\sim 20000$. Therefore
$q > {\mathcal R}^{-1}$ but $q^2 < {\mathcal R}^{-1}$. A gap should form with the gap edge
at the outer Lindblad resonance located, 
for a planet orbiting at $2.5$ au, at a radius $r = 4$ au.
However, for the mass ratio and Reynolds number in this initial simulation, the disc should not 
be strongly truncated at the outer Lindblad resonance. The Lindblad resonance
can be determined from $\Omega(r) = \omega m/(1+m)$ where $\omega$ is the angluar 
frequency of the orbiting planet and, for the outer Lindblad resonance, $m = 1$ 
(Binney \& Tremaine 1987; Lin \& Paploizou 1979a). The
angular frequency at the outer Lindblad resonance is therefore 
$\Omega(r) = \omega / 2$. The assumption of a planet orbiting at $2.5$ au was based on 
SED calculations using an analytic disc geometry and the results of Schneider 
et al.(2002) which suggest that the best fit to the SED occurs for gap extending to a radius of $4$ au
(i.e., the radius of the outer Lindblad resonance for a planet orbiting at $2.5 $ au). 

\begin{figure}
\centerline{\psfig{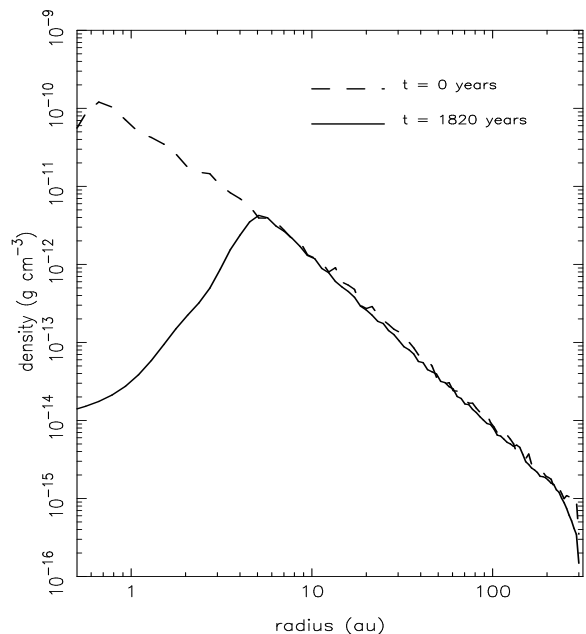}}
\caption{\label{centdens}Azimuthally averaged midplane 
disc density at the beginning (dashed line)
of our simulations and after the $1.7 M_J$ planet has completed $450$ orbits at
$2.5$ au (solid line).}
\end{figure}

Figure \ref{centdens} shows the azimuthally averaged 
midplane density at the beginning (dashed line) and end (solid line) of the 
simulation. After $1820$ years the planet orbiting at $2.5$ au has produced a gap
in the disc out to $r \approx 4$ au. The gap edge is relatively soft, consistent with the relative
magnitudes of the mass ratio ($q$) and Reynolds number (${\mathcal R}$). 
The softness of the gap edge is likely to have an impact on the
radiation transfer through the disc. As will be considered in a later section, to
produce a gap with a significantly harder edge would require a more massive companion
or a significantly less viscous disc. The mass 
within $5$ au has reduced from $4 \times 10^{-4} {M_{\odot}}$ at $ t = 0$ to 
$4 \times 10^{-5} {M_{\odot}}$ at $t = 1820$ years. The reduction in the number of SPH particles respresenting the gas
within $5$ au requires the use of equation \ref{zdens} to
determine the vertical
density at distances greater than $1$ scale height above the disc midplane.

If the planet clears sufficient material from within its orbit, torques
from material outside its orbit should cause it to eventually lose angular
momentum and spiral inwards.  However, the planet in this simulation 
is approximately four times as 
massive as the disc material contained within its orbit and hence we expect it
to suffer minimal orbital migration (Syer \& Clarke 1995; R. Nelson et al. 2000). 

Accretion onto the central star and the orbiting planet continues throughout the
simulation despite the formation of the gap, consistent with simulations performed
by Artymowicz \& Lubow (1996). The central star has increased from it's initial mass
of $0.85 {M_{\odot}}$ to a mass of $0.85019 {M_{\odot}}$ after $1820$ years, while
the orbiting planet has increased from $1.7 \times 10^{-3} M_{\odot}$ to 
$1.88 \times 10^{-3} {M_{\odot}}$ in the same time period. The accretion rate at
the end of the simulation (averaged over the last 100 years)  was 
$\sim 5 \times 10^{-8} M_{\odot} {\rm yr^{-1}}$ and 
$1 \times 10^{-8} M_{\odot} {\rm yr^{-1}}$ for the central star and orbiting planet 
respectively. An accurate 
calculation of 
the accretion rate in our simulation is limited by the mass resolution, since each 
SPH particle has a mass of $1.6 \times 10^{-7} {M_{\odot}}$, and by our simplistic
accretion process. We do not include the rotating magnetosphere of the T Tauri star
which is likely to truncate the disc at the corotation radius and force the accretion
flow to follow the stellar magnetic field lines (Ghosh \& Lamb 1978). The accretion rate from the
model is however similar to that expected for GM~Aur (Gullbring et al. 1998). That
mass is flowing through the gap and accreting onto the central star and planet
is also consistent with other simulations (Artymowicz \& Lubow 1996; 
Lubow et al. 1999). 

The disc velocities also remain very close to Keplerian despite the 
presence of an orbiting
planet that perturbs the potential. This is consistent with CO observations
(Koerner, Sargent \& Beckwith 1993; Dutrey et al. 1998) which suggest that GM~Aur is surrounded
by a Keplerian circumstellar disc inclined at $\sim 50^{\circ}$.

\section{Radiation Transfer Models}

We calculate the SED for our simulations using the Monte Carlo 
radiative equilibrium technique of Bjorkman \& Wood (2001).  The output of
our radiative transfer simulation is the disc temperature structure and the
emergent SED. The disc 
heating is assumed to be from starlight and the Monte Carlo ``photons'' 
are tracked within a two dimensional grid in $r$ and $\theta$.  Our
radiation transfer code calculates disc temperature and does not assume a
vertically isothermal structure as in the SPH simulation. The 
input stellar spectrum is a 4000K, $\log g=3.5$ Kurucz model 
atmosphere (Kurucz 1994) and the stellar radius ($R_*$) is taken to be
1.75 $R_\odot$.  The circumstellar dust opacity is taken to 
be that described in Wood et al. (2002a) which fits the HH~30~IRS SED and 
was also used to fit the GM~Aur SED (Schneider et al. 2002).  The dust model 
has a shallower wavelength dependent opacity than ``standard'' ISM dust 
(e.g., Mathis, Rumpl, \& Nordsieck 1977) and a larger grain size as 
inferred for many other classical T~Tauri stars (e.g., Beckwith et al. 
1990; Beckwith \& Sargent 1991).

We find that the dearth of near-IR excess in GM~Aur's SED cannot be
explained by a dust 
composition different to that used here. Using an ISM mixture and a disc
without a substantial gap still results in a near-IR excess not observed in
GM~Aur. Our results are also in agreement with the Calvet et al. (2002) modelling
of TW~Hya. In their model they do use different dust types in the inner and
outer disc, but still require a very low density inner disc to reproduce
the near-IR emission observed in TW~Hya. Further, ISM grains yield too steep a slope
to the sub-mm SED (see also Beckwith \& Sargent 1991). We therefore conclude that
the dearth of near-IR excess emission from GM~Aur can only be explained by an
evacuated (low density) inner region. 

Circumstellar discs absorb and reprocess stellar photons in a layer 
which, depending on the disc optical depth, can lie four to five 
scaleheights above the disc midplane (D'Alessio et al. 1999).  
Our SPH simulations cannot resolve this low density material, so 
to provide a density grid for our radiation transfer simulations we have 
extrapolated the density, using Equation (\ref{zdens}), to large heights above the disk midplane.  
The density calculated from the SPH simulations was azimuthally averaged and
then interpolated onto 
a two dimensional grid in $r$ and $\theta$.  The grid has 100 cells for 
$0.25{\rm au}\le r\le 300{\rm au}$ 
spaced proportional to $r^3$ and 400 cells linearly spaced for 
$0\le \theta\le \pi$.  This gridding allows us to adequately resolve the 
SPH density for our SED calculations.  

\subsection{GM~Aur SED}
\begin{figure}
\centerline{\psfig{figure=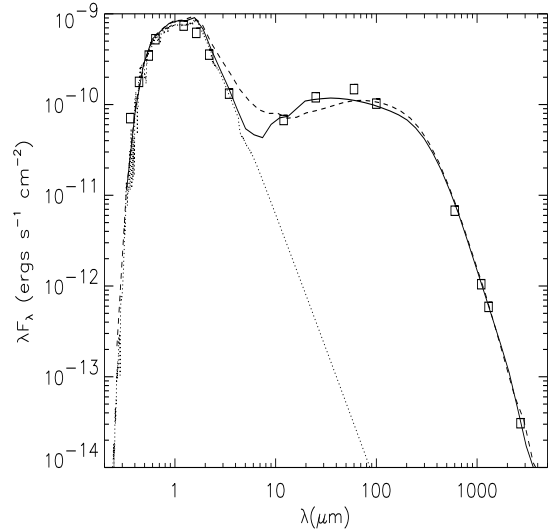,angle=90,width=3.0truein,height=3.0truein}}
\caption{GM~Aur's spectral energy distribution (squares),
our model viewed at $i=50^{\circ}$ (solid line) and input stellar
spectrum (dotted line). The dashed line shows the SED calculated in the absence
of a gap with the disc extending in to $7$ stellar radii. The presence of a gap 
(solid line) clearly reduces the near infra-red flux and enhances the flux at slightly
longer wavelengths.}
\label{sed_2J}
\end{figure}

Figure \ref{sed_2J} shows the GM~Aur SED data. The dotted line shows the input model 
atmosphere, and the open squares are data points obtained using various
instruments (e.g., Cohen \& Kuhi 1979; Kenyon \& Hartmann 1995; Weaver \& Jones 1992;
Beckwith \& Sargent 1991). 
The solid line shows the SED calculated using the azimuthally averaged
SPH density as input to the Monte Carlo radiation transfer calculation. The SPH
simulation was run for $450$ orbits of the $1.7$ Jupiter mass planet at $2.5$ au (1820
years), giving sufficient time for the density profile to reach a steady state. The disc structure
is also approximately azimuthally symmetric since
the viscous timescale is significantly greater than the planet's orbital period. The
inclination of the disc was taken to be $50^{\circ}$, consistent with the value
of $54^{\circ} \pm 5^{\circ}$ determined by Dutrey et al. (1998), and with the
recent scattered light models of Schneider et al. (2002).
The model SED provides a good fit to the observations, particularly 
the lack of near-IR excess and the level and slope of the long wavelength 
continuum. The dashed line in Figure \ref{sed_2J} is the SED for an equivalent disc
in which a gap is not present and in which the inner edge of the disc is taken to be
at $7$ stellar radii. Comparison between the solid line and dashed line clearly shows that
the effect of the disc clearing is to reduce the excess emission at short 
wavelengths ($2 - 10$ microns) and slightly increase the excess at longer wavelengths ($10 - 30$ microns). 

This is consequence of the stellar flux being reprocessed at the disc edge. 
In the simulation with $R_{\rm min} = 7 R_*$ (dashed line
in Figure \ref{sed_2J}) stellar
photons are reprocessed in high temperature ($> 1000$ K) regions of the disc giving a
near-IR excess. When a gap is present the photons are reprocessed further out in the disc
where the temperature is lower, decreasing the near infrared flux and
increasing the flux at longer wavelengths. A fit to the
data is only possible if the gap extends all the way to the inner disc boundary. If the 
planet has only had sufficient time to clear a ring of material, leaving the inner disc
almost unchanged, this inner material should remain hot, resulting in $2$ and $3$ micron
fluxes similar to that for the dashed line in Figure \ref{sed_2J}, and higher than that
observed for GM~Aur (open squares). 

\subsection{Companion (planet) mass and viscosity}
Although we are able to reproduce the observed SED with a $1.7$ Jupiter mass planet 
orbiting at $2.5$ au, the exact form of the density profile depends on the relationship
between the mass ratio, $q$, and the disc Reynolds number, $\mathcal{R}$. A weakness
of SPH is the difficulty in determining exactly the viscosity and hence Reynolds
number.
In light of this, we have performed a number of additional simulations considering
substellar companions (planets) 
of various masses, each of which was run for at least $250$ orbits of
the companion around the central star ($> 1000$ years).
 In these simulations the disc was modelled between
$0.25$ and $100$ au using $150000$ SPH particles with a surface density profile such 
that the total disc mass would be $0.047 M_{\odot}$ were the disc extended to $300$ au. 
Using $150000$ SPH particles to simulate the inner $100$ au of the disc should give 
approximately the same resolution as the $300000$ particles used to simulate
the disc out to $300$ au. The Reynolds number in the simulations using $150000$ 
SPH particles should therefore be very similar to the Reynolds number in a 
$300000$ particle simulation (Murray 1996). The
companion masses considered for these additional simulations 
were $0.085$, $21$, and $43 M_J$. In simulating companions with relatively 
large masses ($21$ and $43 M_J$)
we are not necessarily 
suggesting that one would expect to observe substellar companions with 
such masses, but simply
illustrating how the SED can vary with mass. Strictly speaking, such massive
objects would generally be regarded as brown dwarfs, and we therefore refer to the $21$ and
$43 M_J$ companions as substellar objects rather than planets.

The process of star formation is known to 
preferentially produce binary stellar systems rather than single stars 
(Duquennoy \& Mayor 1991). On the main sequence, however, the mass ratio of close binaries
is rarely so extreme as to pair a Solar mass star with a brown dwarf (Marcy \& Butler 2000;
Halbwachs et al. 2000). If this trend also exists during the pre-main-sequence phase, one 
would expect disc-embedded planetary mass objects to be more common than embedded close
brown dwarfs. The formation mechanism of close binaries is, however, unclear with some
simulations (Bate, Bonnell \& Bromm 2002) suggesting substantial evolution of binary
orbital parameters while a disc is present.  The possibility of observing substellar
mass companions at this early stage could therefore be an important test of different
binary formation models.

The Reynolds number, $\mathcal{R}$, 
in the disc is calculated
to be $\sim 20000$. The $0.085$ and $1.7$ Jupiter mass planets therefore 
have $q > {\mathcal{R}}^{-1}$ but $q^{2} < {\mathcal{R}}^{-1}$ while the $21$ and $43$ Jupiter
mass substellar objects
have $q > {\mathcal{R}}^{-1}$ and $q^{2} > {\mathcal{R}}^{-1}$. The two 
planets ($0.085$ and $1.7 M_J$) 
should therefore produce gaps that are not strongly truncated at the
outer Lindblad resonance while the two substellar objects ($21$ and $43 M_J$)
should truncate the disc strongly 
at the outer Lindblad resonance (Lin \& Papaloizou 1979a; Syer \& Clarke 1995). For a
Reynolds number larger than used here, a planetary mass companion ($< 10 M_J$) 
would satisfy
easily the latter condition ($q > {\mathcal{R}}^{-1}$ and $q^{2} > {\mathcal{R}}^{-1}$). The
results we obtain for the companions referred to as substellar objects would therefore
also be applicable to planetary mass objects in a disc with a Reynolds number larger than in
the simulation presented here. 

\begin{figure}
\centerline{\psfig{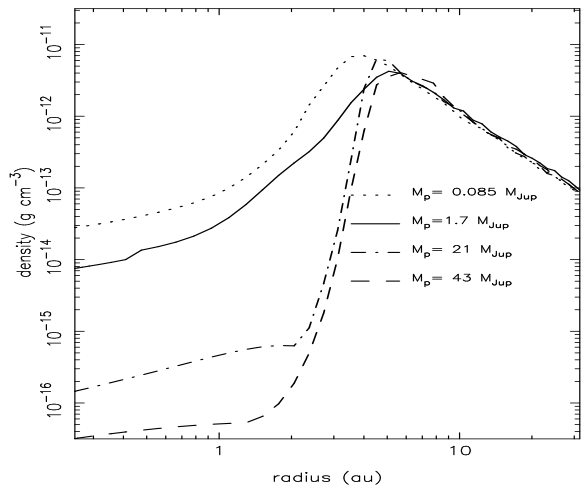}}
\caption{Azimuthally averaged midplane density profiles for substellar objects (planets) 
with masses
of $0.085$ (dotted line), $1.7$ (solid line), $21$ (dash-dot line), and $50$ $M_{J}$
(dashed line). The two higher mass objects truncate the disc more strongly at the
outer Lindblad resonance than the two lower mass planets.}
\label{centdensall}
\end{figure}

Figure \ref{centdensall} shows the azimuthally averaged midplane
densities for the three additional masses together with the azimuthally averaged
midplane density profile for the $1.7 M_J$ 
planet, between radii of $0.25$ and $25$ au. 
As expected the $0.085$ and $1.7$ Jupiter
mass planets produce gaps with soft edges while the $21$ and $43$ Jupiter mass
objects truncate the disc strongly at the outer Lindblad resonance. The position of
the gap edge varies slightly for the different masses, but in all cases is 
close to the outer Lindblad resonance at $\sim 4$ au.

\begin{figure}
\centerline{\psfig{figure=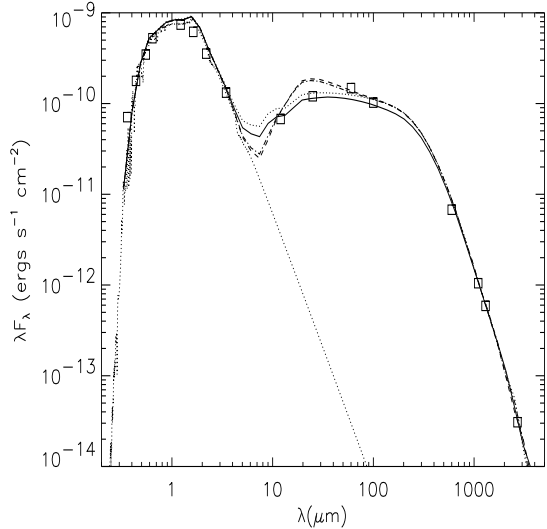,angle=90,width=3.0truein,height=3.0truein}}
\caption{GM~Aur's SED computed using the azimuthally averaged density profiles
due to planets with masses of $0.085$ (dotted line), $1.7$ (solid line) and substellar
companions with masses of $21$ 
(dash-dot line) and $43 M_J$ (dashed line), as input to the Monte Carlo radiation
transfer.}
\label{sedall}
\end{figure}

\begin{figure}
\centerline{\psfig{figure=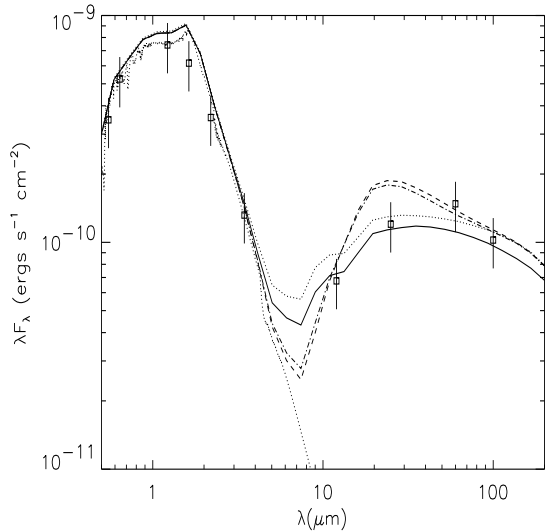,angle=90,width=3.0truein,height=3.0truein}}
\caption{Detailed view of the computed SEDs between $5$ and $200$ microns. The 
line styles correspond to embedded planet masses of $0.085$ (dotted line), $1.7$ (solid
line), $21$ (dash-dot line), and $43 M_J$ (dashed line). Error bars
have been added to the data points with the error ($\pm 25$ \%) based on worst case
error estimates.}
\label{blowup}
\end{figure}

Figure \ref{sedall} shows the SEDs calculated using the azimuthally averaged density
profiles for the four masses as input to the radiation transfer calculation. Since
the simulations considering masses of $0.085$, $21$, and $43 M_J$ were
only performed out to $100$ au, the density profiles were extrapolated to $300$ au.
The solid and dotted lines are for the $1.7 M_J$ and $0.085 M_J$ planets, 
while the dashed and dash-dot lines are for the $43 M_J$ and $21 M_J$ substellar objects.
The SEDs for the $1.7$ and $0.085 M_J$ planets are very similar, as
are the SEDs for the two substellar objects ($21$ and $43 M_J$), 
which are, in fact, very difficult to 
separate. Figure \ref{blowup} shows a detailed view of the computed SEDs 
between $5$ and $200$ microns with the line styles the same as in Figure \ref{sedall}. 
Figure \ref{blowup} also shows $\pm 25$ \% error bars on the
data points (open squares). This choice of error was based on worst case 1 $\sigma$ error estimates
(Cohen \& Kuhi 1979; Weaver \& Jones 1992; Kenyon \& Hartmann 1995). 
Apart from the SED flux at $12$ microns due to the $0.085 M_J$ planet (dotted line) being slightly
higher than that observed, the SEDs computed from the density profiles fit the $2$, $3$, and $12$
micron data points well. The flux between $3$ and $12$ microns varies quite
considerably for the different masses, suggesting that more observations
in this wavelength range could allow for the determination of the companion mass 
or, since the Reynolds number in the disc is unlikely to be accurately known, the
relationship betwen the mass ratio and the Reynolds number. The SEDs due
to the $21$ and $43 M_J$ substellar objects 
also have a flux at $25$ microns that is considerably 
higher than that observed. Although not definitive, this may
suggest that if the GM~Aur SED is due to an embedded companion (planet) 
orbiting at $\sim 2.5$ au, 
the mass must be such as to produce a gap that is not strongly truncated at the
outer Lindblad resonance, i.e. $q > {\mathcal{R}}^{-1}$ but $q^{2} < {\mathcal{R}}^{-1}$. This also
appears to be the case for TW~Hydra as Calvet et al. (2002) require low density material
within 4~au to fit the $2 - 10$ micron SED flux.

At the distance of GM~Aur ($140$ pc), a few Jupiter mass planet orbiting at $2.5$ au will
produce a `wobble' in the central star with an angular displacement of $\sim 0.1$ 
milliarcsec, and a period of $4$ years. Future high precision astrometric instruments
should be sufficiently sensitive to make such an observation. Observations
of this kind have been suggested for detecting both the presence of 
embedded planets (Boss 1998) and
for detecting instabilities in self-gravitating protoplanetary discs (Rice et al. 2002).

\begin{figure}
\centerline{\psfig{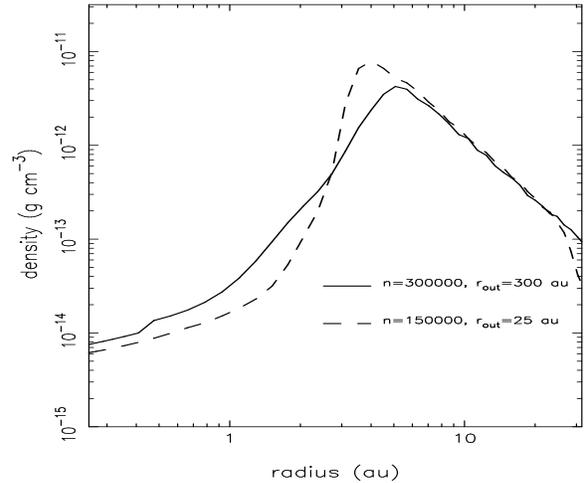}}
\caption{Azimuthally averaged midplane density profiles for a disc modelled, between
$0.25$ and $300$ au using 300000 SPH particles (solid line), and one modelled between
$0.25$ and $25$ au using 150000 SPH particles.  The higher resolution simulation has 
a larger Reynolds number and hence, even though both simulations have the same
planet mass ($1.7 M_J$), the gap is slightly harder than the lower resolution simulation.}
\label{hires}
\end{figure}

Although it is difficult in SPH both to determine and set the viscosity, by 
performing a number of simulations with various companion masses we are able to
determine how the density profile, and resulting SED, varies for various values of
the mass ratio to Reynolds numbers ratio. 
To further test the relationship between planet mass and viscosity we have also
performed a high resolution SPH simulation using 150000 SPH particles to simulate the
inner $25$ au of the accretion disc. This increases the Reynolds number by a factor of
2 and allows the inner few au of the disc to be represented by more SPH particles
once the gap has formed. Figure \ref{hires} shows the midplane disc density for
the simulation using 300000 particles to represent the entire disc (solid line)
and the simulation using 150000 particles to respresent the inner 25 au (dashed
line). In both cases a planet mass of $1.7 M_J$ was used. Although the density
profiles differ, this is entirely consistent with the increased Reynolds number of the
high resolution simulation. Extrapolating the high resolution simulation to 300 au 
and using the resulting density profile as input to the radiation transfer produces
an SED that, like the previous $0.085$ and $1.7 M_J$ simulations, fits the
data remarkably well. 

\section{Summary}

We have presented an SPH simulation of a planet opening up a gap in 
GM~Aur's protoplanetary disk.  For a $1.7 M_J$ planet orbiting at 
2.5~au, the disc density within 4 au is decreased by three orders 
of magnitude.  The almost Keplerian velocity structure 
matches that derived from CO observations.  Further, our radiation 
transfer simulations of a passively heated disk show that the disc
structure from the SPH simulation reproduces the observed SED of GM~Aur.  

Additional simulations of companions (planets) 
with masses of $0.085$, $21$, and $43 M_J$
were also performed.  The structure of the resulting gaps depend on the 
relationship between the mass ratio and the Reynolds number in the disc. For the Reynolds
number in our simulations, the two
planets ($0.085$ and $1.7 M_J$) produce gaps with relatively soft edges,
while the two substellar objects ($21$ and $43 M_J$) truncate the disc strongly at
the outer Lindblad resonance ($\sim 4$ au). As with the $1.7 M_J$ planet, the
SED computed using the density profile due to the $0.085 M_J$ planet fits the data
points extremely well. The SEDs computed using the density profiles due to the two
substellar objects ($21$ and $43 M_J$) 
fit most data points well but have quite a significant excess at $25$
microns. Although not definitive, this may suggest that the mass must be
such as to produce a gap that is not strongly truncated at the outer Lindblad resonance. 
Further observations between $3$ and $20$ microns, where there is a considerable
difference in the computed SEDs, but currently no data points, will be needed to determine
if this is indeed the case. Future SIRTF observations will provide the required measurements
to test our predictions.

Since the density profile depends on the relationship between the mass ratio and the
Reynolds number, additional information would be required to determine the actual planet mass.
A few Jupiter mass planet orbiting at 2.5 au will produce a `wobble' in the central star,
at the distance of GM~Aur ($140$ pc),
with an angular displacement of $\sim 0.1$ milliarcsec, and a period of $4$ years. 
Future high precision astrometric 
instruments should be sufficiently sensitive to make such observations. Not only would this 
establish if a planet is indeed orbiting at $\sim 2.5$ au but should also allow for the 
determination of the planet mass from which the Reynolds
number, and hence the turbulent viscosity, could be estimated.   

The transfer of angular momentum between the embedded planet and the disc material is
ultimately expected to cause the planet to migrate inwards (Lin \& Paploizou 1986).  
Simulations (R. Nelson et al 2000) suggest that the migration 
timescale is of order $10^5$ yrs for a Jupiter mass planet starting at 5 au. 
Since as many as 15 percent of Solar type stars
are expected to have planets within $5$ au, that there are only currently two T Tauri stars
with SED signatures of close-in massive planets is somewhat surprising. This could reflect the
relatively sparse observations in the relevant wavelength range ($3 - 12$ microns). 
Alternatively, fully evacuated inner holes may form in only a fraction of classical T Tauri
stars with planets. An annular gap with a sizeable
inner disc would lead to less obvious signatures in the SED.

\section*{acknowledgements}

We acknowledge financial support from a PPARC standard grant (WKMR) and 
Advanced Fellowship (KW). BW would like to acknowledge support from NASA's Long
Term Space Astrophysics (LTSA) Research Program, NAG 5-8412.

\end{document}